# Multi-objective Distributed Optimization for Zonal Distribution System with Multi-Microgrids

Tao Xu, Lemeng Liang, Zuozheng Liu, Rujing Wang
Key Laboratory of Smart Grid of Ministry of Education,
Tianjin University, Tianjin, China

Lingxu Guo
State Grid Tianjin Electric Power Co.
Tianjin, China

*Abstract*—The issue of voltage variations caused by integration of renewables has been addressed in this paper through distributed management of Microgrids (MGs). The distribution network (DN) takes the network losses and voltage quality as objectives, an alternating direction method of multipliers (ADMM) with adaptive penalty modulation has been proposed to realize distributed optimization with improved convergence. On the basis of satisfying the optimization target on the system level, the individual MG manages the controllable resources inside the MG to minimize the operating cost on the lower level. The feasibility and effectiveness of the proposed method has been demonstrated on a modified IEEE 33-bus system.

## I. Introduction

Distributed energy resources (DERs) have been highly penetrated in distribution networks (DNs), such as photovoltaic (PV), wind power and battery energy storage system (BESS) etc. The traditional operation and dispatching methods of DN are challenged by the wide spread integration of renewables and pressures on carbon reduction [1]. MG, as an effective unit for integrating DERs has been developed rapidly in recent years, clustered MGs have broad application prospects in active distribution networks (ADNs) that have advanced requirements on system operation and control [2]. Hierarchical optimal control and dispatching schemes to accommodate increasing penetration of renewables and operation cost minimization have also attracted significant attention[2],[3].

Comparing to conventional centralized operation schemes, distributed approach can realize local optimization through the coordination among multiple zones of DNs without systemwide information, consequently alleviating the computation burden, congestion and low efficiency. ADMM has been recognized as an important approach to solve the distributed optimization on DNs. Distributed active and reactive power control algorithms on controllable DERs and associate inverters have been developed to solve voltage variations and phase unbalance utilizing ADMM algorithms[4]. A modified ADMM algorithm which could handle non-convex optimization without relaxing the original formulation has been proposed [5]. And an async-ADMM method with iterative varying penalty has been developed to improve its convergence performance [6].

In order to give full play of the regulatory role of MGs in the DN and realize the MG operation cost minimization, a bi-level distributed multi-objective optimization method for DN with multi-MGs has been developed. The upper level aims to minimize the network losses and improve the voltage quality, an ADMM algorithm with improved adaptive penalty modulation has been adopted to realize efficient computation on zonal DNs. Taking the tie-line power optimized by the upper level as a constraint, the lower level tries to achieve the best economic operation by adjusting the controllable DERs inside the MG. The remainder of this paper is organized as follows: Section II presents the zonal partition method. Section III and IV introduce the bi-level optimization model and the ADMM algorithm based on adaptive penalty modulation. Case studies are presented in Section V. Concluding remarks are given in Section VI.

## II. Zonal Partition Method of DNs

### A. Basic concept of zone

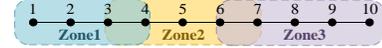

Figure.1 Schematic diagram of zone partition

A typical radial DN can be represented by a graph $G=(N,E)$, where $N$ and $E$ represent the set of nodes and branches of the whole system, $e_{ij} \in E$ represents a branch in the system, $\{i,j\} \in N$. If the DN can be divided into $r$ zones, the set of zones can be represented by $R$. The boundary between adjacent zones is coupled by overlapping branches. Taking a 10 nodes DN shown in Fig.1 as an example, zone1 and zone2 are overlapped through branch $e_{34}$, which can be represented as $O:=\{e_{ij}|E_a \cap E_b, \forall a,b \in R \,\&\, b \neq a\}$. The state variables of $e_{ij}$ include active power $P_{ij}$ and reactive power $Q_{ij}$ transmitted in the branch, and the voltage $V_i$ and $V_j$ of the nodes at both ends of the branch. The state variables of the coupling branch in region $a$ are represented by vectors $X_{a,ij} := \{P_{a,ij}, Q_{a,ij}, V_{a,i}, V_{a,j}\}$.

### B. Voltage sensitivity with MGs

The relationship between the injected power variations of each PQ node and the voltage and phase angle is shown in (1).

$$\begin{bmatrix} \Delta\delta \\ \Delta V \end{bmatrix} = J^{-1} \begin{bmatrix} \Delta P \\ \Delta Q \end{bmatrix} = \begin{bmatrix} \frac{\partial P}{\partial \delta} & \frac{\partial P}{\partial V} \\ \frac{\partial Q}{\partial \delta} & \frac{\partial Q}{\partial V} \end{bmatrix}^{-1} \begin{bmatrix} \Delta P \\ \Delta Q \end{bmatrix} = \begin{bmatrix} \frac{\partial \delta}{\partial P} & \frac{\partial \delta}{\partial Q} \\ \frac{\partial V}{\partial P} & \frac{\partial V}{\partial Q} \end{bmatrix} \begin{bmatrix} \Delta P \\ \Delta Q \end{bmatrix} \quad (1)$$

This work was partly supported by NSFC(51777134, 52061635103)

where, $\Delta P$, $\Delta Q$ are the variations of active and reactive power injected by PQ nodes; $\Delta\delta$ and $\Delta V$ are the changes of voltage phase angle and magnitude of PQ nodes; $J$ is the Jacobian matrix, and the voltage sensitivity of the slack node is 0.

According to (1), the voltage magnitude variations of each PQ node to the injected power changes of each MG connection node can be obtained from $J$ as:

$$[\Delta V] = \left[\frac{\partial V}{\partial P_{MG}} \quad \frac{\partial V}{\partial Q_{MG}}\right]\begin{bmatrix}\Delta P_{MG}\\ \Delta Q_{MG}\end{bmatrix} \quad (2)$$

where, $\Delta P_{MG}$, $\Delta Q_{MG}$ represent the active and reactive power variations of each MG point of common coupling (PCC); $\partial V/\partial P_{MG}$ represents the voltage active power sensitivity matrix of PQ nodes to MG nodes; $\partial V/\partial Q_{MG}$ represents the voltage reactive power sensitivity matrix of PQ nodes to MG nodes.

The larger the corresponding sensitivity value is, the greater the influence of active injection of this MG on PQ node voltage. In order to ensure the MG has the strongest capability to control the nodes in the zone, PQ nodes and their corresponding MG node with the strongest sensitivity are grouped into one zone. The steps of the zone partition are:

Step 1: Obtaining the voltage-active power sensitivity matrix $\partial V/\partial P_{MG}$ of each PQ node to each MG node through $J^{-1}$;

Step 2: Dividing PQ nodes with the highest voltage sensitivity to a certain MG into the same zone;

Step 3: Setting the slack node into the zone it physically connected;

Step 4: Determining the overlapping branches between zones.

### III. BI-LEVEL DISTRIBUTED OPTIMIZATION OF DNs WITH MULTI-MGS

For zonal DN with multi-MGs, a bi-level programming-based optimal scheduling model has been developed in this paper as shown in Fig.2. Aiming at minimum network losses and optimum voltage quality, the distribution network operator (DNO) adopts the distributed optimization method and takes the system power balance as the constraint to schedule the power exchange of PCCs and reactive power of PV inverters at upper level. Taking the tie-line power optimized by the upper level as a constraint, the lower level tries to achieve the best economic operation by adjusting the controllable DERs inside the MG.

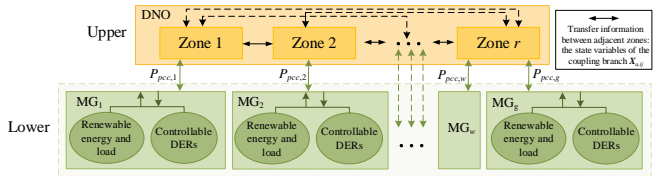

Figure.2 Schematic diagram of DN regulation with multi-MGs

#### A. Optimal dispatching model of DN (Upper level)

A simplified method is adopted to convex the model with the assumptions: 1) power losses on the tie-line is relatively small; 2) voltage deviations between nodes are relatively small.

*1) Multi-objective distributed optimization*

In this paper, a distributed fuzzy optimization method inspired by the principle of fuzzy multi-objective optimization (FMO) is proposed by transforming the multi-objective optimization into a single-objective optimization problem[7]. For a DN with $r$ zones, the concept of regional satisfaction variables is proposed by imitating the principle of classical FMO method. The objective function in zone $a$ is:

$$\begin{cases}\min f_{a1} = \sum_{i=1}^{n_a-1}\sum_{j=i+1}^{n_a} I_{ij}^2 r_{ij} = \sum_{i=1}^{n_a-1}\sum_{j=i+1}^{n_a}(P_{ij}^2+Q_{ij}^2)r_{ij}\\ \min f_{a2} = \sum_{i=1}^{n_a}\left|V_i^2 - V_i^{spec2}\right|\end{cases} \quad (3)$$

For zone $a$, the membership function is:

$$\mu_{as} = \begin{cases}1 & ,f_{as} \leq f_{as}^{\min}\\ \dfrac{f_{as}^{\max}-f_{as}}{f_{as}^{\max}-f_{as}^{\min}}, & f_{as}^{\min} < f_{as} \leq f_{as}^{\max}, s=1,2\\ 0 & ,f_{as} > f_{as}^{\max}\end{cases} \quad (4)$$

where, $\mu_{as}$ represents the membership degree of the $s^{th}$ objective function in zone $a$; $f_{as}^{\min}$ and $f_{as}^{\max}$ are the minimum and maximum of the $s^{th}$ objective function in zone $a$. Let the satisfaction variable $\varphi_a$ of zone $a$ be the minimum of all objective membership degrees, the zonal optimization objective:

$$\begin{cases}\max \varphi_a(x_a)\\ s.t.\begin{cases}\varphi_a \leq \mu_{as}\\ \mu_{as} = \dfrac{f_{as}^{\max}-f_{as}}{f_{as}^{\max}-f_{as}^{\min}}, f_{as}^{\min}<f_{as}\leq f_{as}^{\max}\end{cases}, s=1,2\end{cases} \quad (5)$$

Consequently, the objective function of distributed FMO is:

$$\max \sum_{a=1}^{r}\varphi_a(x_a) \quad (6)$$

*2) Constraints*

Distflow method is utilized to express its constraints [8]:

$$\begin{cases}\sum_{k:j\to k}P_{jk} = P_{ij} - p_j^c + p_j^g\\ \sum_{k:j\to k}Q_{jk} = Q_{ij} - q_j^c + q_j^g\\ U_j = U_i - 2(r_{ij}P_{ij}+x_{ij}Q_{ij})\\ U_1 = V_{ref}^2\end{cases} \quad (7)$$

where: $U_1$ is the square of slack bus voltage with $V_{ref}$ as the reference value, namely $U_j=V_j^2$; $P_{jk}$ and $Q_{jk}$ represent the power receiving from transmission grid; $r_{ij}$ and $x_{ij}$ are the resistance and reactance of $ij$; $p_j^c$ and $q_j^c$ are the active and reactive load of $j$; $p_j^g$ and $q_j^g$ are active and reactive power integrated. The nodal voltages constraints are shown in (8), $\varepsilon$ indicates the allowable voltage deviation:

$$1-\varepsilon \leq V_j \leq 1+\varepsilon \quad (8)$$

The PV normally operates within power factor constraint as:

$$\begin{cases}P_{PV,j} = P_{PV,j}^{MPPT}\\ \cos(\arctan(Q_{PV,j}/P_{PV,j}))\geq 0.95\\ (Q_{PV,j})^2+(P_{PV,j})^2 \leq S_{PV,j}\end{cases} \quad (9)$$

where: $P_{PV,j}$ and $Q_{PV,j}$ are the PV active and reactive power injected by node $j$; $P_{PV,j}^{MPPT}$ indicates the active power injected into node $j$ when PV works at MPPT; $S_{PV,j}$ is the PV installed capacity of node $j$.

And the tie-line need to be managed as:
$$P_{pcc,w}^{\min} \leq P_{pcc,w} \leq P_{pcc,w}^{\max} \quad (10)$$
where, $P_{pcc,w}^{\min}$ and $P_{pcc,w}^{\max}$ are the power thresholds of the tie-line.

### B. Optimal dispatching model of MG (Lower level)

Aiming at the operation cost minimization, the lower level needs to work within the constraints determined by the upper level while maintaining the power balance and accommodating renewables in full.

#### 1) Objective function

The lower level takes the operating cost minimization of each MG as the objective function, i.e microturbine (MT), diesel generator (DE) and BESS, the objective function is:
$$\min f_{MG_w} = C_{MT,w} + C_{DE,w} + C_{BESS,w} \quad (11)$$
where, $C_{MT,w}$, $C_{DE,w}$ are the operating cost of MT and DE in $MG_w$; $C_{BESS,w}$ is the degradation cost of BESS. The cost function of $C_{MT,w}$ and $C_{DE,w}$ are:
$$C_{MT,w} = C_{op,MT,w} + C_{fu,MT,w} \quad (12)$$
$$C_{DE,w} = C_{op,DE,w} + C_{fu,DE,w} \quad (13)$$
where $C_{op,MT,w}$ and $C_{fu,MT,w}$ are the operation and maintenance cost and fuel cost of MT in $MG_w$ respectively, $C_{op,DE,w}$ and $C_{fu,MT,w}$ are the operation and maintenance cost and fuel cost of DE respectively, and their expressions are as follows:
$$\begin{cases} C_{op,MT,w} = K_{op,MT,w} P_{MT,w} \\ C_{fu,MT,w} = K_{fu,MT,w} \cdot V_{fu,MT}(P_{MT,w}) \end{cases} \quad (14)$$
$$\begin{cases} C_{op,DE,w} = K_{op,DE,w} P_{DE,w} \\ C_{fu,DE,w} = K_{fu,DE,w} \cdot V_{fu,DE}(P_{DE,w}) \end{cases} \quad (15)$$
where: $P_{MT,w}$ and $P_{DE,w}$ are the power output of MT and DE in $MG_w$; $K_{op,MT,w}$ and $K_{op,DE,w}$ are the operation cost coefficients of MT and DE, take it as 0.0126\$/kWh and 0.0063\$/kWh; $K_{fu,MT,w}$ and $K_{fu,DE,w}$ are the fuel price of MT and DE; $V_{fu,MT}(P_{MT,w})$ and $V_{fu,DE}(P_{DE,w})$ are the fuel consumption of MT and DE. Assuming the exhaust gas temperature of MT's turbine and bromine cooler remains constant all the time, the fuel consumption is:
$$V_{fu,MT} = \sum P_{MT} \Delta t / (\eta_{MT} \cdot L_g) \quad (16)$$
where, $V_{fu,MT}$ is the consumption of gas; $\Delta t$ is the time interval; $P_{MT}$ is the output of MT; $L_g$ is the low calorific value of natural gas; $\eta_{MT}$ is the power generation efficiency of MT.

The power generation characteristics of DE is:
$$V_{fu,DE} = \alpha (P_{DE})^2 + \beta P_{DE} + \gamma \quad (17)$$
where: $V_{fu,DE}$ is the fuel consumption of DE; $P_{DE}$ is the output of DE; $\alpha$、$\beta$、$\gamma$ are 2.6667, 0.1637 and 0.00015 respectively.

The cost function of $C_{BESS,w}$ is:
$$C_{BESS,w} = P_{BESS,w} \cdot \zeta(P_{BESS,w}) \quad (18)$$
where: $P_{BESS,w}$ is the charging/discharging power of BESS, and $\zeta(P_{BESS,w})$ is the degradation factor, namely 0.123\$/kW.

#### 2) Constraints
Power balance needs to be maintained as:
$$P_{MT} + P_{DE} + P_{BESS} + P_{pcc} = P_{load} - P_{PV} - P_{Wind} \quad (19)$$

where: $P_{MT}$, $P_{DE}$, $P_{BESS}$, $P_{pcc}$, $P_{load}$, $P_{PV}$, $P_{Wind}$ are the power output of MT, DE, BESS, PCC, load, PV and wind in MGs.

MT operation constraints include power and ramping rate:
$$\begin{cases} P_{MT,\min} \leq P_{MT} \leq P_{MT,\max} \\ P_{MT,t} - P_{MT,t-1} \leq R_{MT,up} \Delta t \\ P_{MT,t-1} - P_{MT,t} \leq R_{MT,down} \Delta t \end{cases} \quad (20)$$
where $R_{MT,up}$ and $R_{MT,down}$ are the maximum ramping rate and descent rate of MT. Similarly, DE operation constraints are:
$$\begin{cases} P_{DE,\min} \leq P_{DE} \leq P_{DE,\max} \\ P_{DE,t} - P_{DE,t-1} \leq R_{DE,up} \Delta t \\ P_{DE,t-1} - P_{DE,t} \leq R_{DE,down} \Delta t \end{cases} \quad (21)$$
where $R_{DE,up}$ and $R_{DE,down}$ are the maximum ramping rate and descent rate of DE respectively.

The state of charge (SOC) of BESS need to be managed as:
$$SOC_t = SOC_{t-1} - P_{BESS,t} \Delta t \cdot \eta_{ch} / E_{BESS} \quad (22)$$
$$SOC_t = SOC_{t-1} - P_{BESS,t} \Delta t / \eta_{dis} E_{BESS} \quad (23)$$
$$SOC_{\min} \leq SOC_t \leq SOC_{\max} \quad (24)$$
where: $\eta_{ch}$ and $\eta_{dis}$ are the charging/discharge efficiency of BESS, respectively; $E_{BESS}$ is the capacity of BESS; $SOC_{\min}$ and $SOC_{\max}$ are the limits of SOC with values of 0.2 and 0.9 respectively in this paper.

## IV. IMPROVED ADAPTIVE PENALTY METHOD OF ADMM

### A. Distributed coordination among zones

In order to achieve distributed optimization effectively, the upper level optimization can be equivalently transformed into:
$$\begin{cases} \max \sum_{a=1}^{r} \varphi_a(x_a) \\ s.t. \begin{cases} ha(x_a) = 0 \\ ga(x_a) \geq 0 \\ X_{a,ij} = X_{b,ij}, e_{ij} \in O \end{cases}, a \in R \end{cases} \quad (25)$$
where $h_a(x_a)$ represents the equality constraint of the subproblem of zone $a$; $g_a(x_a)$ represents the inequality constraints of the subproblem; $x_a = \{P_{a,ij}, Q_{a,ij}, U_{a,i}, p^g_{a,i}, q^g_{a,i} \mid \forall i,j \in N_a\}$ represents the set of all variables in zone $a$, and zone $b$ represents the adjacent zone to $a$. Hence, the upper level optimization can be decoupled.

### B. Adaptive penalty modulation of ADMM

The model (25) is a convex optimization model with separable objective functions and linear coupling constraints, so it can be solved in a distributed manner by standard ADMM. Taking two zones as an example:
$$\begin{cases} x_a^{m+1} = \arg\min\left[-\varphi_a(x_a) + \lambda^m(X_{a,ij} - X_{b,ij}^m) + \frac{\rho^m}{2} \|X_{a,ij} - X_{b,ij}^m\|_2^2\right] \\ x_b^{m+1} = \arg\min\left[-\varphi_b(x_b) + \lambda^m(X_{a,ij}^{m+1} - X_{b,ij}) + \frac{\rho^m}{2} \|X_{a,ij}^{m+1} - X_{b,ij}\|_2^2\right] \\ \lambda^{m+1} = \lambda^m + \rho^m(X_{a,ij}^{m+1} - X_{b,ij}^{m+1}) \end{cases} \quad (26)$$

where $m$ is the number of iterations; $\lambda$ is the Lagrange multiplier, $\rho^m$ is the penalty parameter at the $m^{th}$ iteration. In order to improve the computation efficiency of ADMM, an adaptive adjustment mechanism for penalty parameter is introduced.

Taking the first formula of (26) as an example, it can be obtained from its primary and quadratic terms:

$$\lambda^m(X_{a,ij}-X_{b,ij}^m)+\frac{\rho^m}{2}\|X_{a,ij}-X_{b,ij}^m\|_2^2=\frac{\rho^m}{2}\left\|X_{a,ij}-X_{b,ij}^m+\frac{\lambda^m}{\rho^m}\right\|_2^2-\frac{\|\lambda^m\|_2^2}{2\rho^m} \quad (27)$$

Let $u^m=\lambda^m/\rho^m$ and the constant $\|\lambda^m\|_2^2/2\rho^m$ can be omitted, then (26) can be transformed into:

$$\begin{cases} x_a^{m+1}=\arg\min\left[-\varphi_a(x_a)+\frac{\rho^m}{2}\|X_{a,ij}-X_{b,ij}^m+u^m\|_2^2\right] \\ x_b^{m+1}=\arg\min\left[-\varphi_b(x_b)+\frac{\rho^m}{2}\|X_{a,ij}^{m+1}-X_{b,ij}+u^m\|_2^2\right] \\ u^{m+1}=\omega^m u^m+\omega^m(X_{a,ij}^{m+1}-X_{b,ij}^{m+1}) \end{cases} \quad (28)$$

where $\omega^m$ is the correction factor, and its value is $\rho^m/\rho^{m+1}$.

In order to update $u^m$ without the participation of the central coordinator and convert the algorithm into synchronous calculation, the average value of the calculation results of the previous local area and the previous iteration of the adjacent area is selected as the reference values of the next iteration to replace the boundary variable of the adjacent zone:

$$X_{Ka,ij}^m=X_{Kb,ij}^m=\frac{X_{a,ij}^m+X_{b,ij}^m}{2} \quad (29)$$

Substituting it into (28), the iterative solution of region $a$ is:

$$\begin{cases} x_a^{m+1}=\arg\min\left[-\varphi_a(x_a)+\frac{\rho_a^m}{2}\|X_{a,ij}-X_{Ka,ij}^m+u_a^m\|_2^2\right] \\ X_{Ka,ij}^{m+1}=\frac{X_{a,ij}^{m+1}+X_{b,ij}^{m+1}}{2} \\ u_a^{m+1}=\omega_a^m u_a^m+\omega_a^m(X_{a,ij}^{m+1}-X_{Ka,ij}^{m+1}) \end{cases} \quad (30)$$

The iterative convergence for the primal and dual residuals of zone $a$ is shown in (31), $\varepsilon$ is the convergence accuracy.

$$\begin{cases} r_a^m=\|X_{a,ij}^m-X_{b,ij}^m\|_2^2\le\varepsilon \\ d_a^m=\|X_{a,ij}^m-X_{a,ij}^{m-1}\|_2^2\le\varepsilon \end{cases} \quad (31)$$

### C. Improved adaptive penalty method

$r^m$ describes the consistency of coupling variables between adjacent zones, $d^m$ describes the degree of local convergence in the zone. The existing literature usually adopt the following method to modify the penalty parameters of ADMM to balance the convergence speed of primal and dual residuals to zero so as to promote the convergence of $r^m$ and $d^m$:

$$\rho^{m+1}=\begin{cases} \rho^m(1+\lg(d^m/r^m)), & d^m<0.1r^m \\ \rho^m/(1+\lg(r^m/d^m)), & d^m>10r^m \\ \rho^m, & others \end{cases} \quad (32)$$

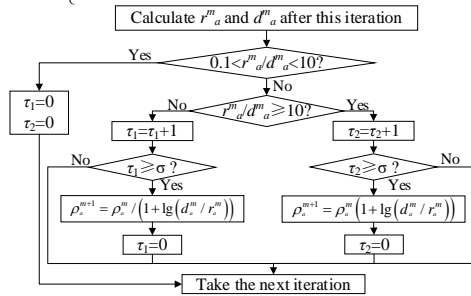

Figure.3 Flow chart of improved adaptive penalty method

The modulation of penalty parameter by conventional methods is too frequent, which may lead to residual oscillation and slow convergence. As shown in Fig.3, an improved penalty parameter modulation method is proposed, the counting factors $\tau_1$, $\tau_2$ and the starting threshold $\sigma$ are determined, the penalty parameter is updated only when the same judgment result is obtained for $\sigma$ consecutive times, which ensures reasonable adjustment on penalty parameter and improves the convergence performance. The overall flow chart is described in Fig.4.

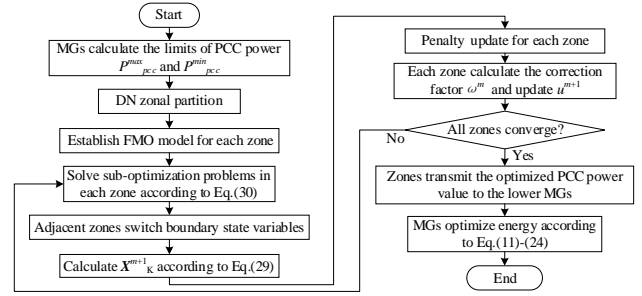

Figure.4 Flow chart of the proposed optimal scheduling model

## V. CASE STUDIES

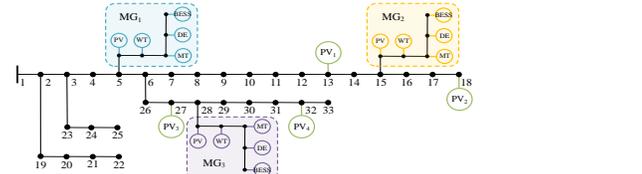

Figure.5 Schematic diagram of case study network with multi-MGs

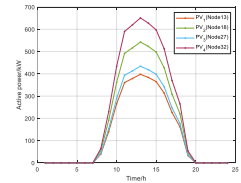 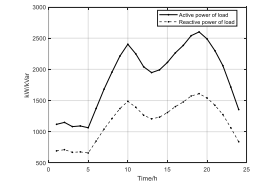

Figure.6 Daily PV profiles    Figure.7 Daily load profiles

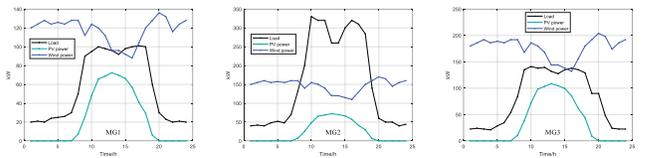

Figure.8 Daily profiles of load, PV and wind in each MG

A modified IEEE 33-bus system as shown in Fig.5 has been utilized for case studies, 3 MGs, 4 PVs (500kW, 700kW, 600kW and 800kW) have been integrated, the associate profiles are shown in Fig.6 and Fig.7. The power output of MT in each MG is between 10~300kW, with initial power 10kW and the maximum ramping rate and descent rate are 120kW and 180kW respectively. The output range of DE is 5~300kW, with initial power 10kW and the maximum ramping rate and descent rate are 160kW and 180kW. The BESS capacity and maximum charging/discharging power of $MG_1$ is 500 kWh and 100kW, and the BESS capacity and maximum charging/discharging power of $MG_2$ and $MG_3$ are 800 kWh and 120kW respectively, the charging/discharging efficiencies are 92% and the initial SOCs are 0.5. The load, PV and wind power profiles of 3 MGs are given in Fig.8. Dynamic gas prices are 0.406\$/m$^3$ at $t$=8-11h, 16-18h; 0.375\$/m$^3$ at $t$=6-7h, 12-15h, 19-21h; and

0.328\$/m$^3$ at $t$=1-5h, 22-24h. The thresholds of voltage are ±0.05p.u. The initial value of penalty $\rho^0$ of ADMM is 250, and the convergence value is $10^{-6}$.

*A. Dynamic zonal results and distributed optimization*

The voltage-active power sensitivity of PQ nodes to each MG is shown in Fig.9, according to the principle of the proposed partition method, the nodes with the same color in the top view are grouped into one zone. It can be obtained intuitively that there are 3 different results during 24h due to the voltage sensitivity variations. In order to verify the feasibility of the proposed method, network losses and nodal average voltage offset before and after optimization can be compared in Fig.10. It can be seen that the network losses and voltage are similar to those obtained by the centralized approach which proves the accuracies. Without multi-MGs' distributed cooperation, the network losses and voltage deviation are large, especially at the peak time, i.e. $t$=20h. The network losses and the voltage deviation can be reduced by 42.69% and 74.83% separately with distributed optimization.

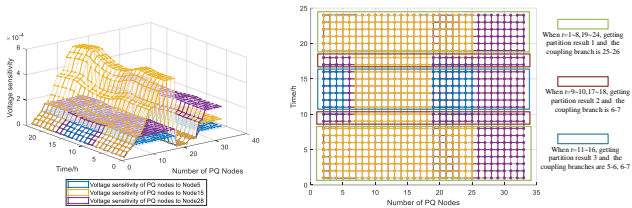

Figure.9 24h voltage-active power sensitivities and zonal results

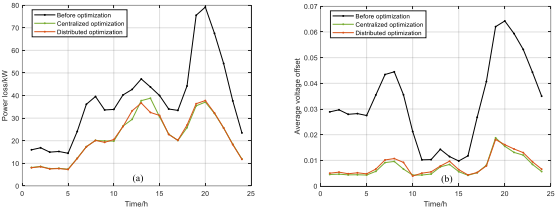

Figure.10 Before/after optimization (a)losses (b) average voltage offset

The reactive power of each PV inverter and power output of the tie-lines are shown in Fig.11-12. It can be seen that the PV inverter can provide reactive compensation, and the 3 MGs balance the fluctuations between supply and demand by optimizing the power flow in a distributed manner.

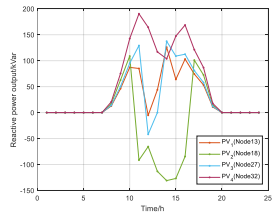 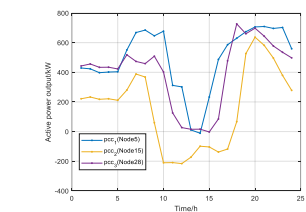

Figure.11 Reactive power output of PV inverters in DN    Figure.12 Power output of the tie-lines

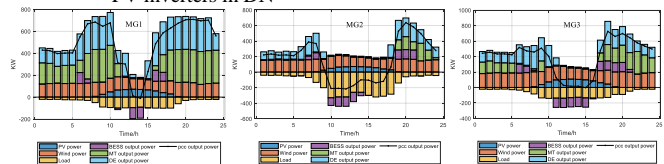

Figure.13 Scheduling results of each MG

The dispatching results of MG1-3 are shown in Fig.13, it worth noting that each MG manages the controllable DERs while meeting the tie-line requirements at the same time without load shedding. In order to meet the tie-line power constraint, the MG schedules the controllable DERs with minimum operation cost, and the BESS acts as the buffer to maintain reliable operation.

*B. Analysis of the improved adaptive penalty method*

The iteration has been compared with traditional methods in Fig.14 to verify the effectiveness of the proposed method. It can be seen that the convergence performance of the traditional adaptive penalty method has been greatly improved compared with fixed penalty factor method, but the number of iterations required for convergence is still relatively high at some points which can be improved by the proposed modulation method, the total number of iterations can be reduced by 60% roughly compared with traditional method. Fig.15 shows the changes of the primal and dual residual of zone a at t=12h, the values are given in logarithmic scale to analyze the variation. With the improved adaptive penalty modulation, the primal and dual residual of each zone can converge at a faster speed and reduce residual oscillation effectively. The improved method can also be applied to large scale distribution system, same performance on calculation efficiency improvement is achieved.

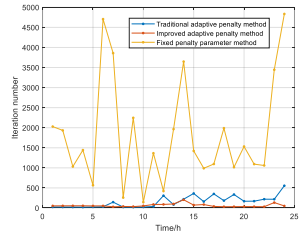 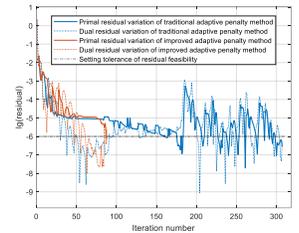

Figure.14 Iteration of ADMM    Figure.15 Primal and dual residuals

## VI. CONCLUSIONS

A distributed FMO method for DN with multi-MGs has been proposed in this paper. On the basis of partitioning the DN according to voltage sensitivity, an ADMM method is proposed to realize distributed optimization with improved convergence performance. The hierarchical optimization model gives full play to the regulation of multi-MGs, and improves the safety and economy of operation while maintaining cost minimization of MG simultaneously. Simulation results have demonstrated the effectiveness and feasibilities of the proposed method.